\documentclass[usenatbib]{aat} 

\usepackage{graphicx}
\usepackage{color}
\usepackage{txfonts}
\usepackage{natbib}
\usepackage{hyperref}
\usepackage{geometry}

\newenvironment{myfont}{\fontfamily{lmtt}\selectfont}{\par}
\DeclareTextFontCommand{\textmyfont}{\myfont}

\def\hb{{\sc{H}}$\beta$\/}

\def\feii{Fe {\sc{ii}}}
\def\caii{Ca {\sc{ii}}}
\def\rfe{R$_{\rm{FeII}}$}
\def\rcat{R$_{\rm{CaT}}$}
\def\cat{CaT}

\def\LLEdd{$L\mathrm{_{bol}}$/$L\mathrm{_{Edd}}$}

\def\RL{$R\mathrm{_{H\beta}}-L_{5100}$}

\def\zsun{Z$_{\odot}$}

\def\un{$\log {U} - \log {n_{H}}$}
\def\rblr{$R\mathrm{_{BLR}}$}
\def\oi{O {\sc i} $\lambda8446$}
\def\hb{H$\beta$}

\def\n{${n_{H}}$}
\def\u{${U}$}

\begin{document}
\setcounter{page}{1}
\nametom{X(X), \pageref{firstpage}--\pageref{lastpage} (XXXX)}
\title{Revisiting the spectral energy distribution of I Zw 1 under the CaFe Project}
\author{Swayamtrupta Panda\inst{1,2}\thanks{CNPq Fellow}, Denimara Dias dos Santos\inst{3}}
\institute{Center for Theoretical Physics, Polish Academy of Sciences, Al. Lotnik\' ow 32/46, 02-668 Warsaw, Poland\\ \email{panda@cft.edu.pl}  
\and
Laborat\'orio Nacional de Astrof\'isica - MCTIC, R. dos Estados Unidos, 154 - Na\c{c}\~oes, Itajub\'a - MG, 37504-364, Brazil
\and
Divisão de Astrofísica, Instituto Nacional de Pesquisas Espaciais, Avenida dos Astronautas 1758, São José dos Campos - SP, 12227-010, Brazil}
\date{Submitted November 2, 2021; Accepted November 6, 2021}
\titlerunning{CaFe Project: I Zw 1 SED}
\authorrunning{S. Panda \& D. Dias dos Santos} 

\maketitle
\label{firstpage}

\begin{abstract} 
The CaFe Project involves the study of the properties of the low ionization emission lines (LILs) pertaining to the broad-line region (BLR) in active galaxies. These emission lines, especially the singly-ionized iron (\feii{}) in the optical and the corresponding singly-ionized calcium (\caii{}) in the near infrared (NIR) are found to show a strong correlation in their emission strengths, i.e. with respect to the broad H$\beta$ emission line, the latter also belonging to the same category of LILs. The origin of this correlation is attributed to the similarity in the physical conditions necessary to emit these lines - especially in terms of the strength of the ionization from the central continuum source and the local number density of available matter in these regions. In this paper, we focus on the issue of the spectral energy distribution (SED) characteristic to a prototypical Type-1 Narrow-line Seyfert galaxy (NLS1) - I Zw 1. We extract the continuum from quasi-simultaneous spectroscopic measurements ranging from the near-UV ($\sim$1200\AA) to the near infrared ($\sim$24000\AA) to construct the SED and supplement it with archival X-ray measurements available for this source. Using the photoionization code \textmyfont{CLOUDY}, we assess and compare the contribution of the prominent ``Big Blue Bump'' seen in our SED versus the SED used in our previous work, wherein the latter was constructed from archival, multi-epoch photometric measurements. Following the prescription from our previous work, we constrain the physical parameter space to optimize the emission from these LILs and discuss the implication of the use of a ``better'' SED.
\end{abstract}

\section{Introduction}
\label{sec1}

The first-ionized state of iron (\feii{}) emission is observed from the ultraviolet to the near infrared (NIR) and acts as one of the main coolants of the broad-line region \citep[BLR,][]{murilo2016,mar18} and manifests as a pseudo-continuum owing to the many, blended multiplets over a wide wavelength range \citep[see][and references therein]{verner99,kovacevic2010}. It is a key parameter in (1) the classification of Type-1 AGNs in the context of the main sequence of quasars \citep{bg92,mar18,panda19b}, and (2) to realize an updated radius-luminosity relation wherein the inclusion of the strength of the \feii{}\footnote{this is well known as the \rfe{} parameter which is the ratio of the integrated \feii{} emission within 4434-4684\AA~ to the broad \hb{} emission.} relates to the accretion rate of the source. Seminal works led by \citet{bg92,verner99,sigut2003} and others encapsulate the `yet to be complete' understanding of the physics of the \feii{} line formation. The \feii{} pseudo-continuum can be modelled appropriately with an 8-dimensional parameter space, encompassing the full diversity of Type-1 AGNs as has been concluded from prior works\footnote{We refer the readers to the PhD Thesis for a comprehensive overview on this issue. A PDF version of the thesis can be accessed using the following \href{https://drive.google.com/file/d/1yNCTNHid2Ev71UwAxCKIILoZLL9C6eFd/view}{link}.}. These 8 parameters consist of the fundamental black hole (BH) and BLR properties, namely (1) the Eddington ratio (\LLEdd{}), (2) the BH mass, (3) the shape of the ionizing continuum\footnote{The shape of the ionizing continuum is a generic term that is used to specify the distribution of the specific photon energy (in units of $\nu$F$_{\nu}$ or $\lambda$F$_{\lambda}$, or in corresponding luminosity units) as a function of frequency ($\nu$) or wavelength ($\lambda$). The term signifies the underlying continuum originating from the very central part of the BH, i.e., the thermally-radiating accretion disk and the Comptonized radiation from the hot/warm corona, that is incident on the BLR cloud.} or the spectral energy distribution (SED), (4) the BLR local density, (5) the metal content in the BLR, (6) the velocity distribution of the BLR including turbulent motion within the BLR cloud\footnote{Mainly with the information of the BH mass and the velocity distribution of the BLR primarily influenced by the central gravitational potential of the BH, and under the assumption of the virial relation, we can derive the distance of the BLR cloud from the BH (i.e. \rblr{}). Thus, the two quantities - the velocity distribution and the \rblr{} are closely connected.}, (7) the orientation of the source (as well as the BLR) with respect to the distant observer, and (8) the sizes of the BLR clouds.

However, the complex electronic structure of \feii{} owing to varied excitation mechanisms makes it difficult to model the atom `perfectly'. This opens up the possibility to search for viable alternatives. Past studies have suggested the existence of a zone shielded from the high-energy photons emanated by the central source and likely located in the outermost portion of the BLR \citep{jol87,dul99,rodriguezardillaetal2002,rodriguez-ardilla2012, rissmann2012,murilo2016} with the presence of emission lines with very low-ionization potentials (IP$\sim$10 eV) such as the \caii{} triplet at $\lambda8498,\lambda8542,\lambda8662$ (hereafter \cat) and \oi{}, in addition to the multiple {permitted} \feii{} transitions. The similarity in the location of the line production of these species suggests a common origin of these LILs, especially between the strengths of the \feii{} and the \cat{}\footnote{these strengths are estimated by normalizing the LILs emission to the broad \hb{} emission and are referred to as \rfe{} and \rcat{}, respectively. Like \rfe{} which is the ratio of the optical \feii{} emission within 4434-4684\AA normalized to the \hb{} emission, the \rcat{} is the ratio of the \cat{} emission normalized also to the same \hb{} emission.}. This has been confirmed both from observational and photoionization studies in our recent works \citep{martinez-aldamaetal15,pandaetal2020_paper1,martinez-aldama_2021}.

The combined importance of the two species has also been recognized in addition to our findings that the \cat{} being an effective proxy serves to be a better alternative to \feii{}-based \RL{} relation \citep{2021POBeo.100..287M}. This is crucial to address the scatter seen due to the inclusion of newer measurements and sources in the \RL{} relation specifically showing a deviation from the classical two-parameter \RL{} relation \citep{bentz13}. A large subset of these sources are noted to belong to the class of Narrow-line Seyfert Type-1 galaxies (NLS1s) that show shorter time delays, and hence a smaller radial distance of the onset of the BLR from the central continuum source (\rblr{}). These smaller \rblr{} values show a marked deviation from the expected \RL{} relation, and addressing this problem is key to our understanding of how these systems evolve and if viable corrections to the classical relation can be made to utilize AGNs as ``standardizable'' cosmological candles.  In addition to this issue, we find in \citet{martinez-aldama_2021} that the ratio of the \cat{} to \feii{} (justifying the project name - CaFe) is an effective tracer of the chemical evolution of AGNs and can help us probe the co-evolution of the AGN and its host galaxy in more detail.

The dearth of observations in the NIR limits our current sample to $\sim$60 sources, but the increased availability of optical and NIR spectroscopic measurements, especially with the advent of the upcoming ground-based 10-metre-class \citep[e.g. Maunakea Spectroscopic Explorer,][]{2019BAAS...51g.126M} and 40 metre-class \citep[e.g. The European Extremely Large Telescope,][]{2015arXiv150104726E} telescopes; and space-based missions such as the James Webb Space Telescope and the Nancy Grace Roman Space Telescope would further help to accentuate the strong correlation shown by these two ionic species.

In this short paper, we focus on a key issue of how the shape of the ionizing SED affects the production of these LILs - especially the \feii{} and \cat{}, and whether it leads to a substantial change in our existing results suggesting a common origin of these species. The paper is organized as follows - In Section \ref{sec2} we outline the photoionization setup and the preparation of the new SED for the prototypical NLS1 - \textmyfont{I Zw 1}. In Section \ref{sec3} we present the results from our analysis and discuss their implication on the existing connection between the two species. We summarize our findings in Section \ref{sec4}.

\section{Methods and Analysis}
\label{sec2}

We apply the photoionization setup prescription that was demonstrated in \citet[][hereafter P21]{panda2021}. We describe briefly the setup here - we perform a suite of \textmyfont{CLOUDY} \citep[version 17.02,][]{f17} models\footnote{N(\u{}) $\times$ N(\n{}) $\times$ N(Z) = 12$\times$11$\times$5 = 660 models}
by varying the mean cloud density, $10^{10.5} \leq n_H \leq 10^{13}\;(\rm{cm^{-3}}$, the ionization parameter, $-4.25 \leq \log U \leq -1.5$, the metallicity, 0.1\zsun{} $\leq$ Z $\leq$ 10\zsun{}, at a base cloud column density, $N_{\rm{H}}$ = $10^{24}$ cm$^{-2}$. The choice for the range for these physical parameters has been studied in detail in prior works \citep{pandaetal2020_paper1,panda2021} especially connected to the low-ionization emission lines (LILs), e.g. \hb{}, \feii{} and \cat{}.

Our main focus in this paper is to highlight the role that the shape of the ionizing continuum (or the spectral energy distribution, i.e. SED) plays in modifying/constraining the physical parameter space for the effective emission of these LILs from the BLR. We assume that the region producing these LILs is dust-free, and the limit on the cloud column density restricts the region to effectively being free of additional scattering effects. In P21 \citep[also in][hereafter P20]{pandaetal2020_paper1}, we presented the results from our photoionization modelling that incorporated a SED for the prototypical NLS1 source \textmyfont{I Zw 1} that was prepared using only photometric data points across a wide wavelength range ($\sim$0.3 mm to $\sim$1200 \AA). We also incorporated in P20 an alternate SED which included photometric data points in the hard X-ray region ($\sim$3 \AA), but this did not change our conclusions in P20 significantly, i.e. the parameter space in terms of the ionization parameter (U) and cloud local density (n$_{\rm H}$) required to maximize the \rfe{} and \rcat{} remain unchanged.

\begin{figure}
\centering
\includegraphics[width=\columnwidth]{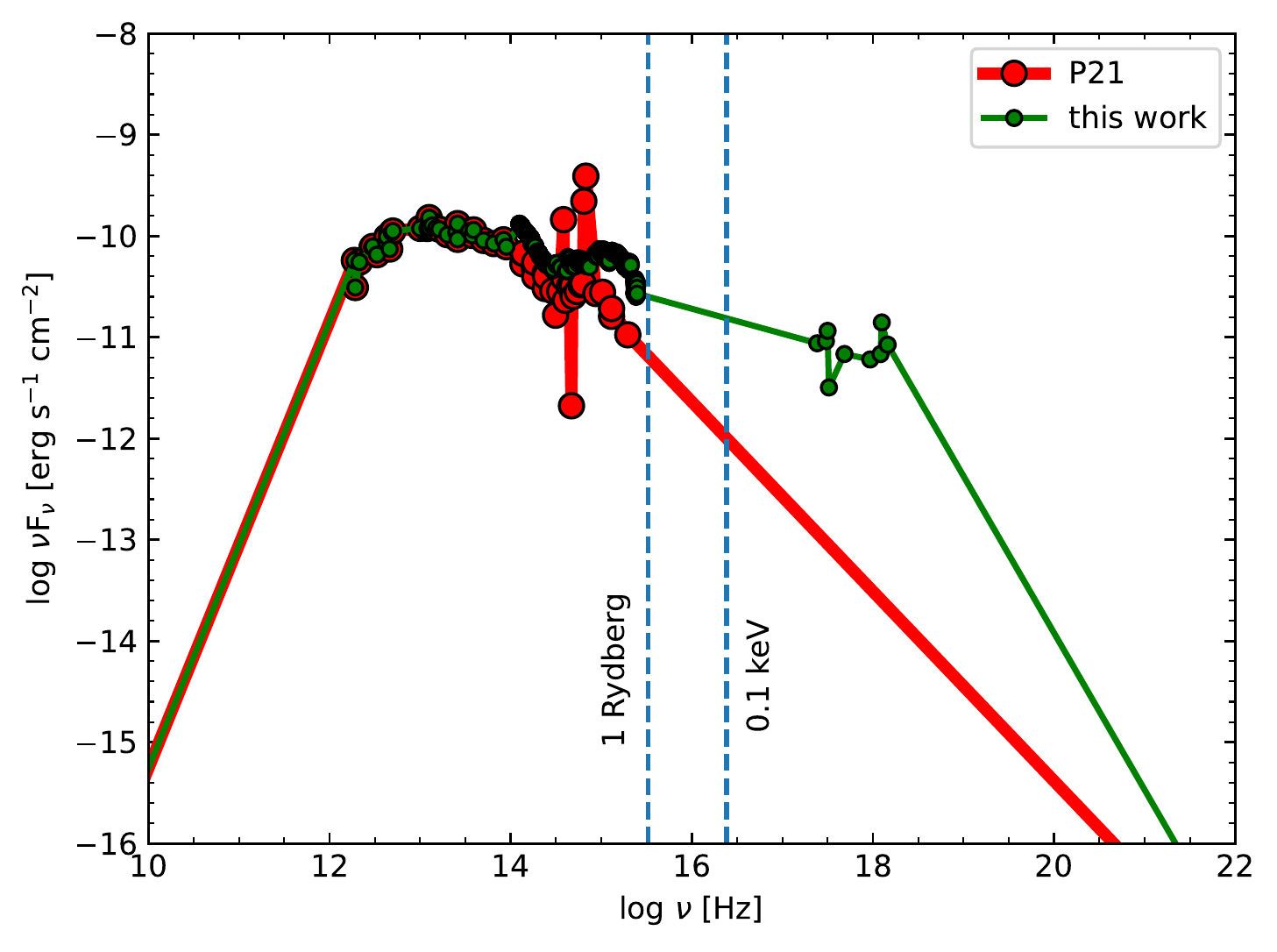} 
\caption{Comparison of the spectral energy distributions (SEDs) for \textmyfont{I Zw 1}. The distribution shown in red represents the original SED considered and the other one (in green) where the SED also includes the X-ray data.}
\label{fig:sed}
\end{figure}

Here, we extract the continuum from the high-resolution composite spectra for \textmyfont{I Zw 1} using archival Hubble Space Telescope (HST) data \citep{bechtold2002} in the UV that is complemented with data in the optical (obtained using the 2.15m Complejo Astronomico El Leoncito - CASLEO) and in the NIR (obtained using the 3.2m NASA Infrared Telescope Facility - IRTF)\footnote{The optical and NIR spectrum were obtained and analyzed in \citet{rodriguezardillaetal2002,riffel2006}.}. For the continuum points extraction, we automatically identify the emission lines, and select regions in the spectrum free of them to extract these points. A full description of the procedure can be found in a different work (Dias dos Santos et al. in prep.). The extracted continuum points are then supplemented with the photometric data points in the X-ray region and wavelengths above 2.5 $\mu$m from the previously used SED in P20 and P21. Figure \ref{fig:sed} shows the comparison between the old SED from P20 and the new SED that is prepared in this work. We can appreciate the ``Big Blue Bump'' feature \citep{c87,panda18b} in our new SED that is more prominent than the older one. This eventually leads to an excess of ionizing photons at the hydrogen ionization limit. Although we are aware that a direct interpolation between the last data point in the UV and the first data point in the X-ray regime is artificial and can lead to an unwarranted excess in the number of ionizing photons, as noticed in our previous work (P20), the hard X-ray photons do not directly interfere in the ionization of the LILs. These hard X-ray photons when coming into contact with the dusty torus (i.e. at distances $\sim$ 1 parsec from the central ionizing source), which has significantly higher optical depths, are known to eventually scatter producing low-energy photons but primarily in the mid to near IR due to the presence of dust \citep{horstetal2008,padovani2017,honig2019}. Thus, these hard X-ray photons have very little role to play in contributing in the ionization of these LILs.

\section{Results and Discussions}
\label{sec3}

\begin{figure*}[!h]
    \centering
    \includegraphics[width=\columnwidth]{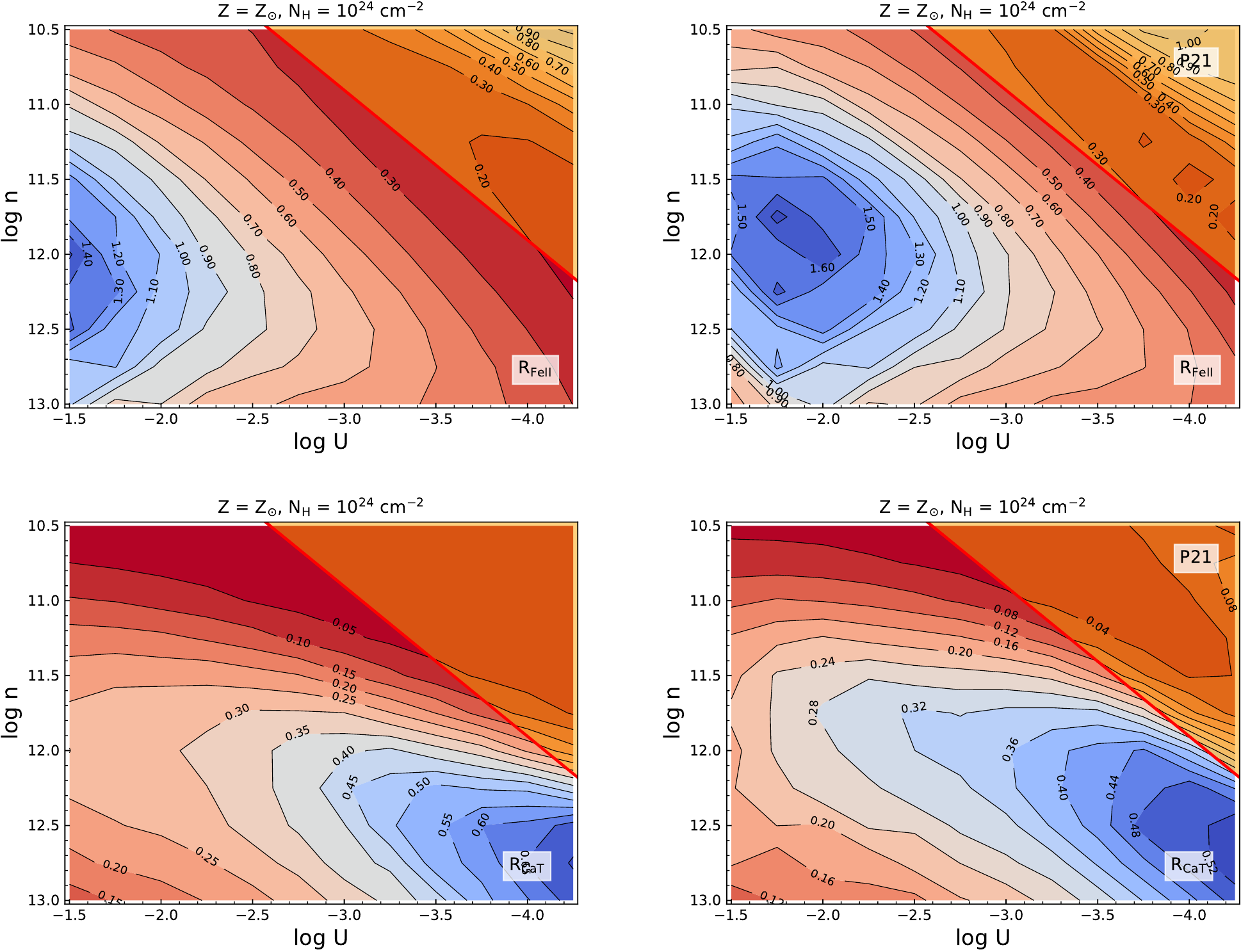}
    \caption{Top panels: \un{} 2D histograms color-weighted by \rfe{} for a characteristic BLR cloud with solar composition (Z=\zsun{}) and column density, N$_{\rm{H}}\,$=$\,10^{24}\,\rm{cm^{-2}}$. Bottom panels: color-weighted by \rcat{}. The left panels incorporate the ``new'' SED (shown in green in Figure \ref{fig:sed}). The right panels are generated using the SED from P20 (shown in red in Figure \ref{fig:sed}). The shaded region in orange represents the dusty region and the onset of the dust (at the dust sublimation radius) is set by the prescription of \citet{Nenkova2008} and shown with the red line.}
    \label{fig2}
\end{figure*}

\begin{figure*}[!h]
    \centering
    \includegraphics[width=\columnwidth]{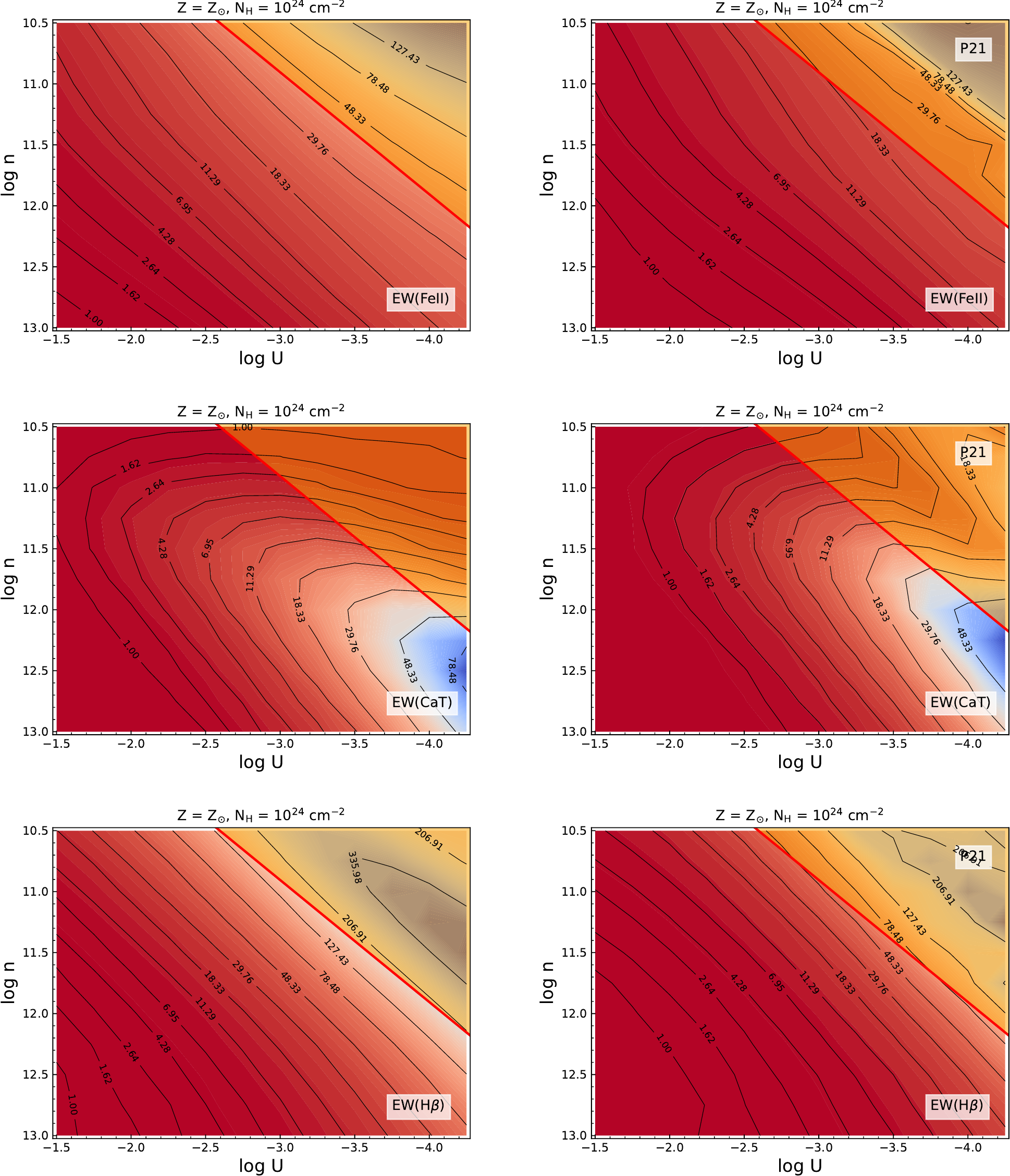}
    \caption{Top panels: \un{} 2D histograms color-weighted by the EW (in units of \AA) of optical \feii{}, middle panels: color-weighted by the EW of \cat{}, and, bottom panels: color-weighted by the EW of \hb{}. The left panels incorporate the ``new'' SED (shown in green in Figure \ref{fig:sed}). The right panels are generated using the SED from P20 (shown in red in Figure \ref{fig:sed}). Other parameters are identical to Figure \ref{fig2}.}
    \label{fig3}
\end{figure*}

Figure \ref{fig2} shows the \un{} parameter space for the \rfe{} (upper panels) and \rcat{} (lower panels). The diagnostic plots on the left row are obtained with the new SED while those on the right are for the older SED (used in P20 and P21). We incorporate the prescription from \citet{Nenkova2008} to separate the dusty and non-dusty regime in the BLR, which has a form: 
\begin{equation}
\rm{R_{sub}} $=$ 0.4\left(\frac{L_{\rm UV}}{10^{45}}\right)^{0.5},    
\end{equation}
where, $\rm{R_{sub}}$ is the sublimation radius (in parsecs) computed from the source luminosity that is consistent for a characteristic dust temperature. This is a simplified version of the actual relation which, in addition to the source luminosity term, contains the dependence on the dust sublimation temperature and the dust grain size. We assume a dust temperature $T_{sub}$ = 1500 K, which has been found consistent with the adopted mixture of the silicate and graphite dust grains, and a typical dust grain size, a=0.05 microns. The dependence of the $\rm{R_{sub}}$ on the temperature is quite small - the exponent on the temperature term is -2.8. On the other hand, the dust grain size is a more complex problem, yet the value adopted is fair in reproducing the characteristic dust sublimation radius in our case \citep[see][for more details]{Nenkova2008,honig2019}. The sublimation radius, hence, is estimated using only the integrated optical-UV luminosity for \textmyfont{I Zw 1}. This optical-UV luminosity is the manifestation for an accretion disk emission and can be used as an approximate for the source's bolometric luminosity. The bolometric luminosity\footnote{This bolometric luminosity value is quite similar to the value obtained by integrating the area under the curve in our new SED.} of \textmyfont{I Zw 1} is L$_{\rm{bol}} \sim 4.32\times10^{45}$ erg s$^{-1}$. This is obtained by applying the bolometric correction prescription from \citet{netzer2019} to \textmyfont{I Zw 1}'s optical monochromatic luminosity, $\rm{L_{5100}} \sim 3.48\times 10^{44}\;\rm{erg\;s^{-1}}$ \citep{persson1988}. This uniquely sets the dust sublimation radius at $\sim 0.83 \rm{pc}\; $(= $ 2.56\times 10^{18}\;$cm). Projecting this sublimation radius on the \un{} plane allows us to recover the non-dusty region that well represents the physical parameter space consistent with the emission from the BLR. This dust-filtering is applied to the models in a post-photoionization stage.

Focusing first on the upper panels in Figure \ref{fig2}, the \rfe{} plots show a slight change in the location of the maximum, the new diagnostics suggest an ionization parameter that is about 0.25 dex higher and a shift by a similar factor is noticed in the local density, albeit reduced.  Also, the value of the \rfe{} obtained at the maximum is reduced by 20\%. To assess the radial distances of the \feii{} emitting region, we use the formulation from P21, i.e.: 

\begin{equation}
    R_{BLR} $=$ \sqrt{\frac{Q_H}{4\pi Un_Hc}} \equiv \sqrt{\frac{L_{bol}}{4\pi h\nu\ Un_Hc}} \eqsim \frac{2.294\times 10^{22}}{\sqrt{Un_H}} 
    \label{eq1}
\end{equation}

where, \rblr{} is the distance of the emitting cloud (in cm) from the ionizing source which has a mean local density, \n{}, and receives an ionizing flux that is quantified by the ionization parameter, \textit{U}. $Q_H$ is the number of ionizing photons, which can be equivalently expressed in terms of the bolometric luminosity of the source per unit energy of a single photon, i.e. h$\nu$. Here, we consider the average photon energy, h$\nu$ = 1 Rydberg \citep{wandel99, marz15}. The specific value of the bolometric luminosity corresponds to \textmyfont{I Zw 1}. Comparing the two plots in the upper panel (\rfe{} based), the maximum \rfe{} emitting location is shifted inwards by a factor 2 (for the P21-based SED plot, the maximum \rfe{} is obtained for an ionization parameter, log \textit{U} = -1.75, at a local cloud density, \n{} = 10$^{11.75}$ cm$^{-3}$. This returns a value for the \rblr{} = 2.294$\times$10$^{17}$ cm. On the other hand, for the new SED, the maximum \rfe{} is obtained for a log \textit{U} = -1.5, at \n{} = 10$^{12}$ cm$^{-3}$, which gives a \rblr{} = 1.294$\times$10$^{17}$ cm). 

For the \rcat{}, the location of the maximum value remains unchanged, although there is a slight increase in the net value in \rcat{} with the new SED. The corresponding \rblr{} location based on the maximum \rcat{} location, is estimated to be about a factor 10 larger than the \rblr{} for maximum \rfe{}. We comment on this result in the next paragraphs.

As was inferred in P21, the location on the \un{} plane that leads to the maximum value for the flux ratios (\rfe{} or \rcat{}) do not agree in terms of their line equivalent widths when compared with observed estimates. As noticed through our simulations, for example, if the ratio of the EW(\feii{}) to EW(\hb{}) is taken (i.e., the \rfe{}), we can notice that at the location of the maximum value for \rfe{}, the EW(\feii{}) is about 4\AA, while for the EW(\hb{}) at that same location the value obtained is about 2.5-3\AA. This is in contradiction to the observed EWs measured from spectral fitting. In addition, such low EWs are almost at the limit of (or below) the observed spectral resolution. The regions wherein we find agreement both in terms of the flux ratios and the corresponding line EWs for these LILs are shifted towards lower ionization parameters (log U $\sim$ -3.0 and lower) for both these lines. We show the corresponding equivalent widths plots in Figure \ref{fig3}. The equivalent widths for \feii{} and \hb{} have been estimated using the continuum luminosity very close to the 5100\AA\ (at 4885.36\AA) and assuming a covering fraction of 20\%, a value consistent with previous studies \citep{baldwin2004,2001ApJ...553..695K,sarkar2020,panda2021}. For the \cat{}, we utilize a continuum closer to the line, i.e. 8329.68\AA, and assume the same covering fraction. As can be noticed in the panels for these LILs in Figure \ref{fig3}, the EWs agreeable to observed estimates ($\sim$30-40\AA) suggest a lowering in the log \textit{U}, below -3.0, for the \feii{} emission. A similar shift is required for the \cat{} emission. This has already been noticed in P21 that at solar composition (without any microturbulence effects), the requested \rfe{} and \rcat{} values cannot be retrieved without agreeable EWs for these lines. An increase in the metal content (up to a factor 3-10) is required to match the observed flux ratios and line EWs for these LILs in \textmyfont{I Zw 1}. The similarities thus obtained in the \un{} parameter space brings the location of the emitting regions for the two species - \feii{} and \cat{}, almost to similar values of \rblr{}.

When compared with the observed flux ratios, we notice that the solar composition models shown here are insufficient to reproduce the \rfe{} estimate by \citet{persson1988}, i.e. 1.778$\pm$0.050 or the more recent estimate by \citet{murilo2016}, i.e. 2.286$\pm$0.199. As was concluded in P21 and also confirmed in \citet{sniegowska21}, there is a need to increase the metal content to super-solar values, i.e. 3-10 \zsun{}, to push the \feii{} emission and thus the \rfe{} estimate in perfect agreement with both these observed estimates. In contrary, the \rcat{} estimates are successfully reproduced in these models with solar composition. \rcat{} estimates for \textmyfont{I Zw 1} are reported in Table 1 in P21: 0.513$\pm$0.130 \citep{persson1988} and 0.564$\pm$0.080 \citep{murilo2016}.

\begin{figure}[!h]
    \centering
    \includegraphics[width=0.5\columnwidth]{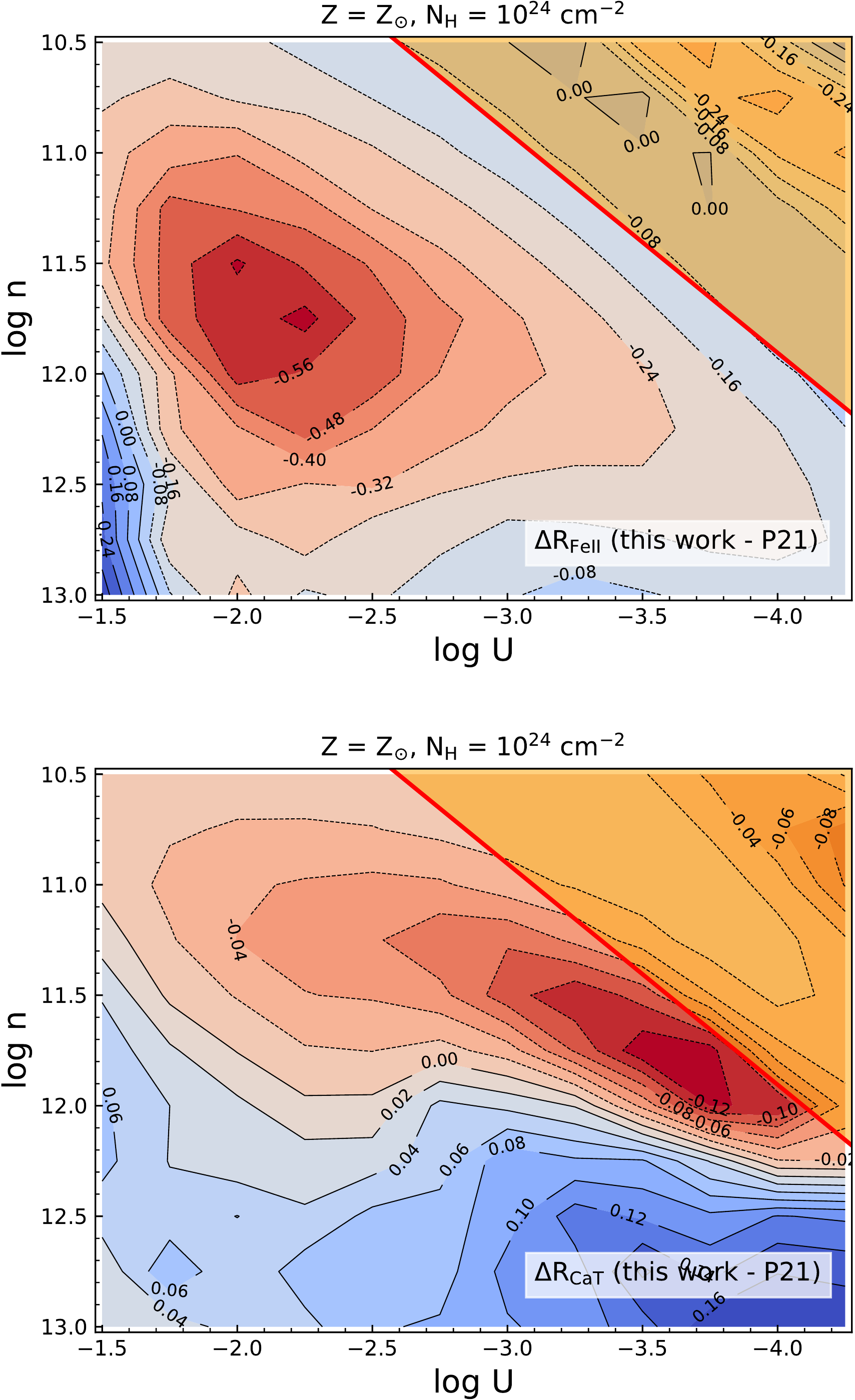}
    \caption{Top: \un{} 2D histograms color-weighted by the difference between the \rfe{} obtained using the two SEDs shown in Figure \ref{fig:sed}. Bottom: color-weighted by the difference between the \rcat{}. Other parameters are identical to Figure \ref{fig2}.}
    \label{figA1}
\end{figure}

\begin{figure}
    \centering
    \includegraphics[width=0.5\columnwidth]{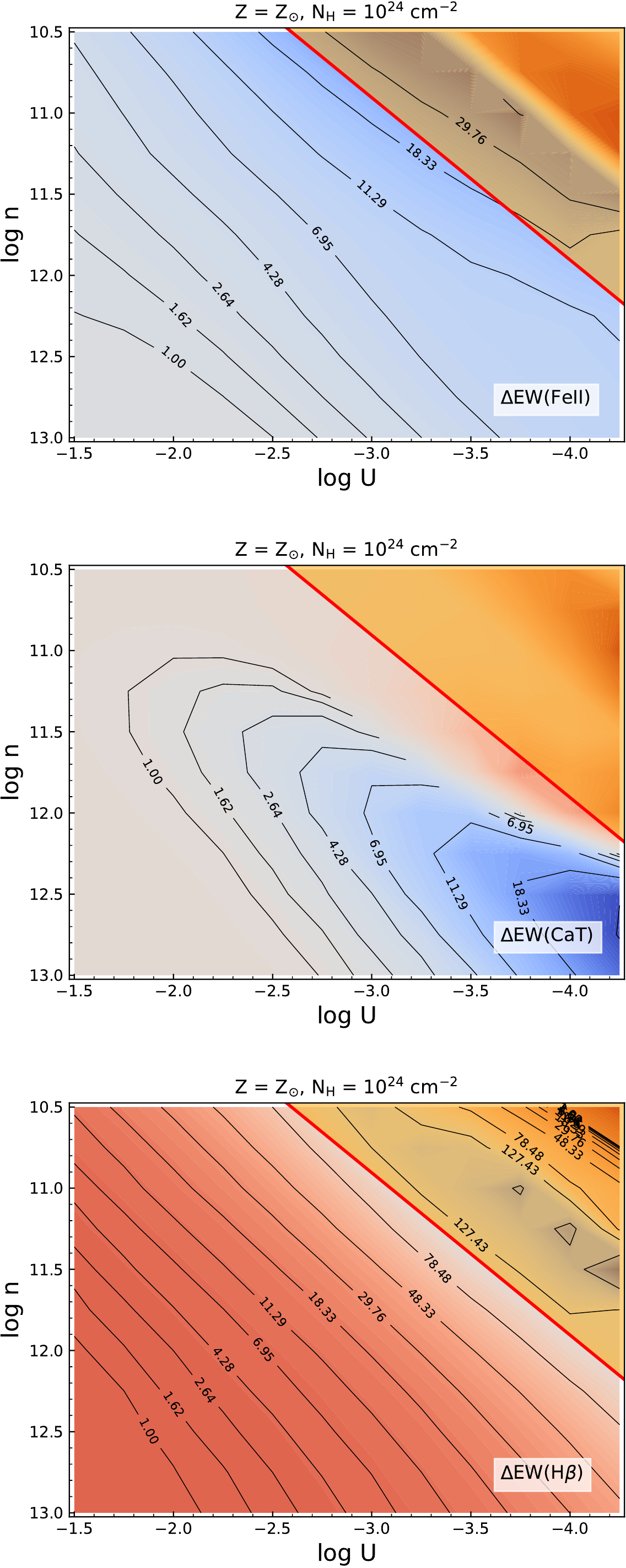}
    \caption{Top: \un{} 2D histograms color-weighted by the difference between the EW of optical \feii{} obtained using the two SEDs shown in Figure \ref{fig:sed}. Middle: color-weighted by the difference between the EW of \cat{}. Bottom: color-weighted by the difference between the EW of \hb{}. Other parameters are identical to Figure \ref{fig2}.}
    \label{figA2}
\end{figure}

To highlight the salient differences in incorporating the new SED in place of the existing SED from P21 for \textmyfont{I Zw 1} we show $\Delta$\rfe{} and $\Delta$\rcat{} plots for our models in Figure \ref{figA1}. Locating the solutions in the \un{} plane where we get agreeable EWs for the two species, we notice that the new SED leads to a lower \rfe{} compared to the other SED - the new SED predicts a \rfe{} value which is lower by about -0.3. But, for the \rcat{}, we retrieve an estimate that is only slightly higher than the previous estimate from P21, i.e. higher by about 0.02-0.04 that is well within the scatter in the observed flux ratio obtained by \citet{persson1988} and \citet{murilo2016}. Thus, we can conclude that there is not a significant change in the predicted \rfe{} and \rcat{} (and their corresponding line EWs) with the incorporation of a ``better'' SED, and hence, the conclusions obtained from our earlier analysis in P21 remain valid. In addition, we show the change in the line EWs in Figure \ref{figA2}, wherein we can notice that for the location in the \un{} plane with the optimal values for EWs, i.e. $\sim$40\AA\ for \feii{} and $\sim$100\AA\ for \hb{} \citep[see Table A1 in][]{martinez-aldama_2021}, which corresponds to a log \textit{U} $\lesssim$ -3.0 around BLR densities log \n{} $\sim$ 10$^{12}$ cm$^{-3}$, we have a $\Delta$\feii{} around 10-20\AA, while for \hb{} this difference rises to be around 30-50\AA. This leads to the slump in the \rfe{} that we notice in Figure \ref{figA1}. While in the case of the $\Delta$\cat{}, there is only a marginal change in the EW in the same region, i.e. about 5-10\AA, which confirms the almost no change in the \rcat{} plots. Thus, the EWs for these LILs provide a better insight to the changes arising due to the change in the SED - an increase in the prominence leads to an increase in the line EWs for these LILs, while the ratios (\rfe{} and \rcat{}) remain rather unaffected.

\section{Conclusions and Future Work}
\label{sec4}

In this paper, we focus on the issue of the spectral energy distribution (SED) characteristic to a prototypical Type-1 Narrow-line Seyfert galaxy (NLS1) - I Zw 1. We extract the continuum from quasi-simultaneous spectroscopic measurements ranging from the near-UV ($\sim$1200\AA) to the near infrared ($\sim$24000\AA) to construct the SED and supplement it with archival X-ray measurements available for this source. Using the photoionization code \textmyfont{CLOUDY}, we assess and compare the contribution of the prominent ``Big Blue Bump'' seen in our SED versus the SED used in our previous work, wherein the latter was constructed from archival, multi-epoch photometric measurements. Following the prescription from our previous work, we constrain the physical parameter space to optimize the emission from these low-ionization lines (LILs) and discuss the implication of the use of a ``better'' SED. We find:

\begin{itemize}
    \item There is only a very slight difference in the estimated flux ratios, i.e. \rfe{} and \rcat{} when we replace the SED from the one used in our previous works \citep{pandaetal2020_paper1,panda2021} to a new, better one which is made by extracting the continuum from spectra for \textmyfont{I Zw 1} ranging from the UV to the NIR.\\
    \item The inclusion of the X-ray data in the construction of the new SED doesn't lead to any significant change in the retrieved flux ratios (or their corresponding line EWs).\\
    \item The BLR clouds need to be selectively overabundant in iron to reproduce the observed \feii{} emission, i.e. up to 3-10 times the solar values. On the contrary, the \cat{} emission predicted from our models agrees with the observed values.\\
    \item The analysis presented here and in P21 highlight the importance to consider the comparison of the line EWs, in addition to just the flux ratios, which leads to a significant improvement to break the degeneracy and exclusion of imposter solutions. The EWs for these LILs provide a better insight to the changes arising due to the change in the SED - an increase in the prominence leads to an increase in the line EWs for these LILs, while the ratios (\rfe{} and \rcat{}) remain rather unaffected.\\
    \item We are successful in constraining the physical parameter space to optimize the emission from these low-ionization lines originating from the BLR and re-affirm the similarity in the location leading to the emission of the \feii{} and \cat{} emission.
\end{itemize}

Our conclusions obtained suggest the importance of constructing better SEDs and utilizing them instead of generic ones. In this work, we focused on the study and analysis of the emission lines pertaining to the BLR which are governed by the ionizing photon flux especially from the energy range that corresponds to the radiation from the accretion disk and the Comptonized corona. We show that the continuum extraction from the broad-band spectrum for \textmyfont{I Zw 1} is able to highlight the prominence of the Big Blue Bump feature in the SED. This was not accounted for in the earlier works and led to interesting results, especially in the line EWs recovery where we see a marked increase in the recovered EWs compared to our previous works. There is still progress needed to fully account for the X-ray radiation, i.e. having a self-consistent SED where we account for the continuum in the X-ray and construct an SED where we also model the region where we have the galactic absorption. This can be tested with currently available \textmyfont{xspec} models, e.g. OPTXAGN \citep{done12}. Also, there is a need to test our findings for other sources similar to \textmyfont{I Zw 1}. In addition to these, there is a need to test and account for the changes in the accretion disk structure (and the corona) as a function of the increasing accretion rate. This increase in the accretion rate has been shown to modify the standard \citet{ss73} disk to a more intricate, slim disk \citep{abramowicz88} solution. Such a change can lead to the development of a inner funnel in the very inner regions of the disk one that is the closest to the BH. This can modify significantly the overall photon energy distribution and make the radiation from these regions more anisotropic. BLR clouds which are located closer to the disk surface would preferably see a continuum that contain less energetic while the distant observer may see a very different SED, more rich in high energy photons coming from the region around the inner-most stable circular orbit of the BH. This anisotropy can help quantify the difference in the SED seen by the BLR clouds in comparison to the distant observer \citep[see][for a qualitative overview]{panda2021}. This work is also in progress and will be presented in a forthcoming work.

\begin{acknowledgements}
   I would like to thank Prof. Bo\.zena Czerny, Prof. Paola Marziani and Dr Mary Loli Mart\'inez-Aldama for fruitful discussions, and to Dr Murilo Marinello and Prof. Alberto Rodr\'iguez-Ardila for assisting with the extraction of the I Zw 1 continuum. The project was partially supported by the Polish Funding Agency National Science Centre, project 2017/26/\-A/ST9/\-00756 (MAESTRO  9), MNiSW grant DIR/WK/2018/12 and acknowledges partial support from CNPq Fellowship (164753/2020-6).
\end{acknowledgements}

\section*{Softwares}
\textmyfont{CLOUDY} v17.02 (\citealt{f17}); \textmyfont{MATPLOTLIB}  (\citealt{hunter07}); \textmyfont{NUMPY} (\citealt{numpy})

\bibliographystyle{aat.bst}
\bibliography{references}


\label{lastpage}
\end{document}